\documentclass{article}
\usepackage{epsfig}
\usepackage{amssymb}
\usepackage{graphicx}
\usepackage{amsmath}
\usepackage{amsfonts}

\date{\today}

\begin{document}

 \title{ 
The Gregory-Laflamme instability and non-uniform 
\\
generalizations of NUT strings
} 

\author{{\large Burkhard Kleihaus, Jutta Kunz} 
{\large } 
and {\large Eugen Radu}  \\ 
 {\small  Institut f\"ur Physik, Universit\"at Oldenburg, Postfach 2503
D-26111 Oldenburg, Germany} }

\newcommand{\vphi}{\varphi}
\newcommand{\vepsilon}{\varepsilon}
\newcommand{\DS}{\displaystyle}
\newcommand{\pih}{\frac{\pi}{2}}
\newcommand{\sqdetg}{\sqrt{-g}}
\newcommand{\sqdetgi}{\frac{1}{\sqrt{-g}}}
\newcommand{\edil}{e^{2\kappa \phi}}
\newcommand{\bA}{\bar{A}}
\newcommand{\bF}{\bar{F}}
\newcommand{\bD}{\bar{D}}
\newcommand{\bx}{\bar{x}}
\newcommand{\beq}{\begin{equation}}
\newcommand{\eeq}{\end{equation}}
\newcommand{\beqs}{\begin{eqnarray}}
\newcommand{\eeqs}{\end{eqnarray}}
\newcommand{\reef}[1]{(\ref{#1})}
\newcommand{\tr}{\mbox{\rm tr}}
\newcommand{\ra}{\rightarrow} 
\newcommand{\be}{\begin{equation}}
\newcommand{\ee}{\end{equation}}
\newcommand{\bea}{\begin{eqnarray}}
\newcommand{\eea}{\end{eqnarray}}
\newcommand{\Ord}[2]{\mathcal O \left(#1\right)^{#2}}

\maketitle

 \begin{abstract}
We explore via linearized perturbation theory the Gregory-Laflamme instability of the  NUT string 
($i.e.$ the $D=4$ Lorentzian NUT solution uplifted to five dimensions). 
Our results indicate that the Gregory-Laflamme instability persists in the presence of a NUT charge $n$,
the critical length of the extra-dimension increasing with $n$ for the same value of mass.
The non-uniform branch of  NUT strings is numerically extended into the full nonlinear
regime.
\end{abstract} 

%%%%%%%%%%%%%%%%%%%%%%%%%%%%%%%%%%%%%%%%%%%%%%%%%%%%%%%%%%%%%%%%%%
\section{Introduction}
%%%%%%%%%%%%%%%%%%%%%%%%%%%%%%%%%%%%%%%%%%%%%%%%%%%%%%%%%%%%%%%%%%

It is well known that any solution of  vacuum general relativity in 
$D$-spacetime dimensions can be promoted to a solution of the same theory in $(D+p)$-spacetime dimensions 
by adding a number of $p$ flat directions. 
These extra-dimensions are usually supposed to be compact, the model being described by a Kaluza-Klein theory.
The simplest such configuration is found by trivially extending to $(D+1)$-spacetime dimensions  
the Schwarzschild black hole in $D$-dimensions.
The resulting solution corresponds to a uniform black string,
the extra-dimension being periodic with an (arbitrary) length $L$. 

Rather unexpectedly, twenty years ago Gregory and Laflamme (GL) made the  discovery that, for a given value of $L$,
the Schwarzschild black string is classically unstable against linearised gravitational
perturbations  below a critical value of the mass
\cite{Gregory:1993vy}.  
Following this discovery, a branch of non-uniform black string (NUBS)
solutions breaking the translational invariance along the periodic direction 
was found perturbatively from the critical
GL string  
\cite{Gubser:2001ac}, 
\cite{Wiseman:2002zc}, 
\cite{Sorkin:2004qq}. 
This non-uniform branch was subsequently numerically extended into the full nonlinear regime in  
\cite{Wiseman:2002zc}, 
\cite{Kleihaus:2006ee}, 
\cite{Sorkin:2006wp}.
Further developments have proven that the GL instability is a generic property of black objects
in spacetimes with compact extra dimensions.
%and became a fruitful field of research.
This includes also
rotating solutions \cite{Kleihaus:2007dg},
 non-vacuum solutions 
\cite{Gregory:1994bj},
\cite{Kleihaus:2009ff},
\cite{Frolov:2009jr}
and configurations with several extra-dimensions
compactified on a torus
\cite{Kol:2004pn} 
(see 
\cite{Kol:2004ww},
\cite{Harmark:2007md},
\cite{Obers:2008pj}
 for reviews of these aspects).

However, most of the work on the stability and phases of black strings has been performed  assuming
that the  solutions approach at infinity the Minkowski spacetime  
times a circle.
Then it is worth inquiring, what happens if we drop these asumptions? 
Will the GL instability persist? 
As proven in \cite{Brihaye:2007ju}, 
this is the case in the presence of a negative cosmological constant,
since the 
anti-de Sitter black strings  
\cite{Copsey:2006br},
\cite{Mann:2006yi}
are also unstable for small enough values of the event horizon radius.

%\newpage

In some sense, at least for $D=4$,
the minimal deviation from the asymptotic flatness 
is to include a ``dual" or ``magnetic" mass in the theory. 
As explicitly proven by the famous Taub-NUT solution 
\cite{Taub:1950ez},
\cite{NUT},
\cite{Misner},
general relativity  
allows for 
``gravitational dyon'' 
solutions  possessing both ordinary (or ``electric'') and ``magnetic'' mass (the NUT charge).
In this case, the metric is not asymptotically flat in the usual sense
although it does obey the required fall-off conditions.
The Taub-NUT spacetime has a number of unusual properties,
  becoming renowned for being 
{\it ``a counter-example to
almost anything}" 
\cite{misner-book}  
and is unlikely to be of interest as a model for a macroscopic object. 
Nevertheless, the Euclideanized  Taub-NUT solution
extremizes
 the gravitational action functional and
might play an important role in the context
of quantum gravity \cite{Hawking:ig},
% extremizing
% the gravitational action functional in Euclidean quantum gravity 
providing an analogue of instantons in gauge theories.

Let us mention also that the vacuum  Taub-NUT solution has been generalized in different directions,
by including matter fields or a cosmological constant \cite{kramer}.
There are also some indications that the NUT charge is an important ingredient
in low energy string theory (see $e.g.$ \cite{Johnson:2004zq}).
However, the pathological features of the vacuum  Taub-NUT solution
are generic and
affect gravitational solutions with ``dual" mass in general 
\cite{Magnon}. 

Of interest in this work is the fact that, being Ricci flat, the Lorentzian Taub-NUT 
solution can be promoted to a solution of the
$D=5$ vacuum Einstein equations.
However, due to the presence of a NUT charge, the asymptotics  
are different from the case of a Schwarzschild black string.
Then it is interesting to inquire if the NUT strings are also unstable.
The main purpose of this paper is to answer this question.
In addition, since all solutions are found to possess
a GL unstable mode, we construct numerically 
the corresponding branch of non-uniform 
configurations.

The  paper is structured as follows: in the next Section we 
review the basic properties 
of the $D=4$ Lorentzian signature Taub-NUT solution. 
The $D=5$ uniform string whose stability is investigated
is the  product of this four dimensional
 solution and a circle.
In Section 3  we investigate its stability and show results for the GL mode obtained 
by numerical calculation.  
The basic properties of the non-perturbative
$D=5$ configurations with a dependence of
the extra-dimension are discussed in Section 4.
We conclude with Section 5 where the results are compiled.

%%%%%%%%%%%%%%%%%%%%%%%%%%%%%%%%%%%%%%%%%%%%%%%%%%%%%%%%%%%%%%%%%%
\section{ The $D=4$ NUT spacetime}
%%%%%%%%%%%%%%%%%%%%%%%%%%%%%%%%%%%%%%%%%%%%%%%%%%%%%%%%%%%%%%%%%%

The line element of the $D=4$ Taub-NUT spacetime  is usually written as
\begin{eqnarray}
\label{TN}
ds^2=\frac{dr^2}{f(r)}+g(r)(d\theta^{2}+\sin^{2} \theta  d\varphi^{2})
-f(r)(dt+2 n \cos \theta d\varphi)^{2}.
\end{eqnarray}
where
\begin{eqnarray}
\label{f}
f(r)=1-\frac{2(m r+n^2)}{r^2+n^2},~~~g(r)=r^2+n^2.
\end{eqnarray}
%where $\theta$ and $\varphi$ are spherical coordinates on $S^2$, with the usual range.
This spacetime has two independent parameters, $m$ and $n$, corresponding to the ``electric"
and ``magnetic" masses, respectively.
The NUT charge $n$ plays a dual role to ordinary  mass, 
in the same way that electric
and magnetic charges are dual within Maxwell theory \cite{dam}.
The Killing symmetries of the  solution are still translations 
and $SO(3)$ rotations.
However, the spherical symmetry in a conventional sense is lost   
when the NUT parameter is nonzero, since the 
rotations act on the time coordinate as well.

This solution has an outer horizon located at
\begin{eqnarray}
\label{rH}
 r_H=m+\sqrt{m^2+n^2}>0.
\end{eqnarray} 
Here $f(r_H) = 0$ is only a coordinate singularity where all curvature invariants are finite. 
A nonsingular extension across this null surface can be found just as at the event horizon of a black hole.
However, the line-element (\ref{TN}) possesses also
an inner horizon at $r_{i}=m-\sqrt{m^2+n^2}<0$.
The solution derived by Taub in 1951 is valid in the "inner" region
with $f(r)<0$, being interpreted as
a cosmological model.
The metric valid in the outer region 
$r\geq r_H$ (which is of interest in this work),
has been derived independently in 1961 by Newman, Unti and Tamburino,
being usually called the NUT space–time\footnote{As discussed by Misner in \cite{Misner}, the NUT space–time   can be 
joined analytically to the Taub space–time
as a single Taub-NUT space–time.}.

One can easily see that, for $n\neq 0$,
the metric (\ref{TN}) has a singular symmetry axis (defined by $\theta=0,\pi$).
As discussed in \cite{misner-book},
these singularities can be removed by appropriate identifications and changes in 
the topology of the spacetime manifold, which imply a periodic time coordinate\footnote{This comes essentially from the fact that
the nondiagonal part of the metric (\ref{TN}) can be generalized to 
$g_{\varphi t}=-f(r) (2n\cos \theta+n_0)$, with $n_0$ an arbitrary constant,
see $e.g.$ the discussion in 
\cite{Manko:2005nm},
\cite{Kagramanova:2010bk}.}. 
However, in this work, following Ref. \cite{Manko:2005nm},
we choose a different interpretation of (\ref{TN}), with 
$-\infty<t<\infty$ and
 two physical  singularities at $\theta=0$ and $\theta=\pi$, respectively.
These singularities are interpreted as two semi-infinite
counter-rotating rods.
Note that the pathology of closed timelike curves 
is still present in this case,
as proven by the fact that, for any $\theta$, the metric component
$g_{\varphi \varphi}$
becomes negative for $r<r_c$,
with $r_c$ a solution of the equation
$\cos 2\theta=\frac{r_c^2+n^2-4n^2 f(r_c)}{r_c^2+n^2+4n^2 f(r_c)}$
\cite{Bonnor1},
\cite{Bonnor2}.
As a result, similar to the case of G\"odel's rotating universe\footnote{This analogy becomes 
more transparent after noticing that the  G\"odel-type universe
corresponds to the
boundary metric of the Taub-NUT solution with a negative cosmological constant
\cite{Radu:2001jq},
\cite{Astefanesei:2004kn},
\cite{Brihaye:2009dm}.} 
\cite{Godel:1949ga},
the Killing vector $\partial/\partial \varphi$ is timelike
in a region around the symmetry axis (which extends to infinity for $\theta \to 0,\pi)$,
and the spacetime 
is not globally hyperbolic.

This interpretation of the solution
leads, however, to an interesting analogy between the 
angular momentum of the NUT charged spacetimes
in general relativity
and that of the spinning solitons of the Georgi-Glashow model.
To this aim, following \cite{Astefanesei:2006zd},
we compute the mass and angular momentum of the 
NUT solution by employing the quasilocal formalism in conjuction with the boundary counterterm method,
which avoids the choice of a reference background.
In this approach one supplements
the gravity action (which contains the Gibbons-Hawking boundary term \cite{Gibbons:1976ue})
  by including suitable boundary counterterms, which are functionals
of curvature invariants of the induced metric on the boundary.
The usual choice
\cite{Lau:1999dp}
%\cite{Mann:1999pc}
for the boundary countertem is 
% \begin{eqnarray}
%\label{Ict}
$
I_{ct}=-\frac{1}{8 \pi G}\int_{\partial {\cal M}} d^{3} x \sqrt{-h}\sqrt{2 \mathcal{R}},
$
%\Psi(\mathcal{R}),
%\end{eqnarray}
where $\mathcal{R}$ is the Ricci scalar of the induced metric on the
boundary $h_{ij}$.
%and
%\begin{eqnarray}
%\label{Ict-n}
%\Psi(\mathcal{R})=
%\sqrt{2 \mathcal{R}}.
%\end{eqnarray}
Varying the total action %(which contains the Gibbons-Hawking boundary term)
with respect to the
boundary metric $h_{ab}$, we obtain the divergence-free boundary stress-tensor
%
%\begin{eqnarray}
%\label{Tik}
$
T_{ab}=
%\frac{2}{\sqrt{-h}}\frac{\delta I}{\delta h^{ab}}=
\frac{1}{8\pi G}\Big( K_{ab}-h_{ab}K
- \Psi (\mathcal{R}_{ab} - \mathcal{R} h_{ab} )
- h_{ab}\Box \Psi
+\Psi_{;ab}
\Big),
$
%\end{eqnarray} 
where $K_{ab}$ is the extrinsic curvature of the boundary and $\Psi=\sqrt{2/\mathcal{R}}$.
Provided the boundary geometry has an isometry generated by a
Killing vector $\xi ^{i}$, a conserved charge
%
%\begin{eqnarray}
$
{\frak Q}_{\xi }=\oint_{\Sigma }d^{2}S^{i}~\xi^{j}T_{ij} 
$
%\label{charge}
%\end{eqnarray}
%
can be associated with a closed surface $\Sigma $. 

Similar to the case of a Schwarzschild black hole,
the boundary of the NUT spacetime is taken at constant $r$, being  sent to infinity
in the final relations.
A straightforward computation gives
$8 \pi G T_t^t= {2m}/{r^2}+O(1/r^3)$,
which 
leads to the usual expression of the ``electric" mass
(which is the charge associated with the Killing vector $\partial/\partial t$)
\begin{eqnarray}
\label{mass-TN}
M=\frac{m}{G}=\frac{1}{G}\frac{r_H^2-n^2}{2r_H}.
\end{eqnarray}
Interestingly, a gravitational dyon possesses a nonvanishing angular momentum density,  
\begin{eqnarray}
\label{J-TN}
8 \pi G T_{\varphi}^t=\frac{4 m n \cos \theta }{r^2}+O(1/r^3).
\end{eqnarray} 
However, one can easily verify that 
the total angular momentum $J$ 
(which is the charge associated with the Killing vector $\partial/\partial \varphi$) 
vanishes.
Noticing that $T_{\varphi t}$ is antisymmetric with respect
 to a reflection in the equatorial plane,
 one can say that a NUT spacetime consists of two counter-rotating regions,
which  agrees with the results in \cite{Manko:2005nm}.
 
The same quasilocal approach applied to the Kerr black hole 
leads to the usual expressions for the conserved charges \cite{Astefanesei:2006zd}.
However, as discussed in 
\cite{Gross:1983hb},
\cite{Manko:2009xx},
\cite{Argurio:2009xr},
one may think of the Kerr metric as possessing also 
a NUT dipole in addition to the usual ``electric" mass.
%
%Thus, we conclude that the general relativity NUT ``dyon" does not rotate globally,
% while the angular momentum is nonzero for the Kerr solution
% which has a vanishing net ``magnetic" mass.
 Thus we note that the simplest NUT ``dyon" in
general relativity  does not rotate globally,
whereas the angular momentum is nonzero for the Kerr solution,
which possesses a vanishing net ``magnetic" mass.
 This reveals an interesting analogy with 
the spinning dyons and dipoles in the Georgi-Glashow model,
featuring an $SU(2)$ gauge field and a Higgs field in the  
adjoint representation.
This model possesses globally regular, particle-like solutions,
the  BPS monopole \cite{'tHooft:1974qc}
and dyon \cite{Julia:1975ff}
being the best known examples.
The existence of a profound connection between the angular momentum and
the electric and magnetic charges 
in this theory has been suggested
already in the seminal paper \cite{Julia:1975ff}. 
Indeed, as discussed in 
\cite{VanderBij:2001nm},
\cite{Kleihaus:2005fs} 
the total angular momentum of the solitons endowed with a 
net magnetic charge vanishes (despite the fact that their angular
momentum density can be nonzero). 
At the same time, the 
angular momentum of a spinning magnetic dipole is nonzero \cite{Paturyan:2004ps}.

Returning to the properties of NUT charged spacetimes,
we notice that  
their thermodynamical description is still poorly understood
(the difficulties result mainly from the absence 
of a global Cauchy surface).
Most of the existing results in the literature
were found by using a  Euclidean approach\footnote{An interesting result
here is that the entropy of such solutions generically does
not obey the simple "quarter-area law", see the discussion in \cite{Astefanesei:2004kn}.}.
The Euclidean version of (\ref{TN}) is obtained 
 by performing the  analytic continuation $n\to iN$, $t\to i \tau$;
 then the regularity of the metric
fixes $m$ as a function of $N$ \cite{Hawking:ig}.
However,  the relevance of the 
results found  on the Euclidean section 
for the Lorentzian signature solution is unclear \cite{Kerner:2006vu}.
Nevertheless, one can define a temperature of solutions
via the surface gravity associated with the Killing vector $\partial/\partial t$
\begin{eqnarray}
\label{TH-TN}
T_H=\frac{\kappa}{2\pi}=\frac{1}{4 \pi r_H}~,
\end{eqnarray}
and also an (outer) event horizon area
\begin{eqnarray}
\label{AH-TN}
A_H=4\pi (r_H^2+n^2).
\end{eqnarray}
Let us also mention that,
although one can write a Smarr-type formula \cite{Holzegel:2006gn}, 
the issue of the first law of thermodynamics for NUT-charged
solutions is unclear.

We close this part by remarking that,
different from the case of a Schwarzschild black hole,
 a negative value  of the electric mass $m$ is 
allowed for the NUT solution. 
Such configurations have $0<r_H<n$ and do not possess a Schwarzschild limit.
Also, since the line-element (\ref{TN}) is invariant 
under the transformation $n\to -n$,
one can take $n>0$  without loss of generality 

%%%%%%%%%%%%%%%%%%%%%%%%%%%%%%%%%%%%%%%%%%%%%%%%%%%%%%%%%%%%%%%%%%
\section{The Gregory-Laflamme instability}
%%%%%%%%%%%%%%%%%%%%%%%%%%%%%%%%%%%%%%%%%%%%%%%%%%%%%%%%%%%%%%%%%%

The line element (\ref{TN}) can trivially be extended to $D=5$
by adding a $dz^2$ term to that metric, 
the extra-coordinate $z$ possessing a
periodicity $L$.
It is natural to expect that the resulting uniform solution becomes unstable at critical values
of $M,n$.
To determine these critical values, we follow  the same approach as in \cite{Gubser:2001ac}.
The starting point is to consider the 
following ansatz for the non-uniform 
generalization of the NUT string\footnote{Similar to the case of a $D=4$ NUT spacetime, 
the singularities at $\theta=0,\pi$ can be eliminated by a coordinate transformation together with a  periodic
identification of $t$.} 
\begin{eqnarray}
\label{metric-NUBS}
ds^2=e^{2B(r,z)}(\frac{dr^2}{f(r)}+dz^2)+g(r)e^{2C(r,z)}(d\theta^{2}+\sin^{2} \theta  d\varphi^{2})
 -f(r)e^{2A(r,z)}(dt+2 n \cos \theta  d\varphi)^{2},
\end{eqnarray}
the uniform limit  corresponding to $A = B = C = 0$.
The problem is thus characterized by two dimensionless parameters:
%\begin{eqnarray}
%\label{ls} 
$\mu_1= M G/{L^{2}}$
and 
$\mu_2= {n}/{L}$.
%\end{eqnarray} 
Here $M$ is the mass of the $D=5$ solutions, as computed from the relation (\ref{2}) below. 
(Note that the uniform solutions have $M=Lm/G$.)
The limit $\mu_2\to 0$ corresponds to the Schwarzschild black string solution in a Kaluza-Klein theory,
in which case 
the GL unstable mode occurs for
$\mu_1 \simeq 0.0649519$ \cite{Gregory:1993vy}.

In the next step, we perform an expansion of the functions $A,B,C$ in terms 
of a small parameter $\epsilon$ and consider a Fourier series in the $z$ coordinate. 
In leading order, we assume:
\begin{eqnarray}
X(r,z) = \epsilon X_1(r) \cos(k z) + O(\epsilon^{2}),
\label{Xseries}
\end{eqnarray}
$X$ denoting generically $A,B,C$ and $k$ being the critical wavenumber corresponding to a static perturbation,
$k= 2 \pi/L$.
This expansion is appropriate for studying perturbations at the wavelength which is marginally stable.

We then substitute the form \eqref{metric-NUBS} in the general Einstein equations and  
expand $A,B,C$ according to \eqref{Xseries}.
The system relevant for addressing the stability problem
is found  by taking  the linear terms in the infinitesimal parameter $\epsilon$. 
Similarly to the $n=0$ case \cite{Gubser:2001ac},
the  Einstein equation $G_r^z=0$
allows to eliminate the function $B_1$ in favor of the other functions
 and to reduce the problem to a system of two differential equations for $A_1$ and $C_1$.
 These equations read:
\begin{eqnarray}
\nonumber
&&
A_1''+\frac{1}{2}(\frac{3f'}{f}
+\frac{2g'}{g})A'
+\frac{f'}{f}C_1'
+\frac{4n^2}{g^2}(A_1-2C_1)
+\frac{8n^2f}{g(gf'+2fg')}(A_1'+2C_1'+\frac{f'}{2f}A_1+\frac{g'}{g}C_1)
-k^2\frac{A_1}{f}=0, 
\\
\label{eqs-p}
&&
C_1''
+ (\frac{f'}{f}+\frac{2g'}{g})C_1'
+\frac{g'}{2g}A_1'
+\frac{4n^2}{g^2}(C_1-A_1)
-\frac{2(2n^2f+g)}{g(gf'+2fg')}(4 C_1'+2A_1'-\frac{f'}{f}C_1+\frac{f'}{f}A_1)
-k^2\frac{C_1}{f}=0,~~~{~~~}
\end{eqnarray}

%%%%%%%%%%%%%%%%%%%%%%%%%%%%%%%%%%%%%%%%%%%%%%%%%%%%%%%%%%%%%%%%%%%%%%%%%%%%%% 
\setlength{\unitlength}{1cm}
\begin{picture}(8,6) 
\put(-0.5,0.0){\epsfig{file=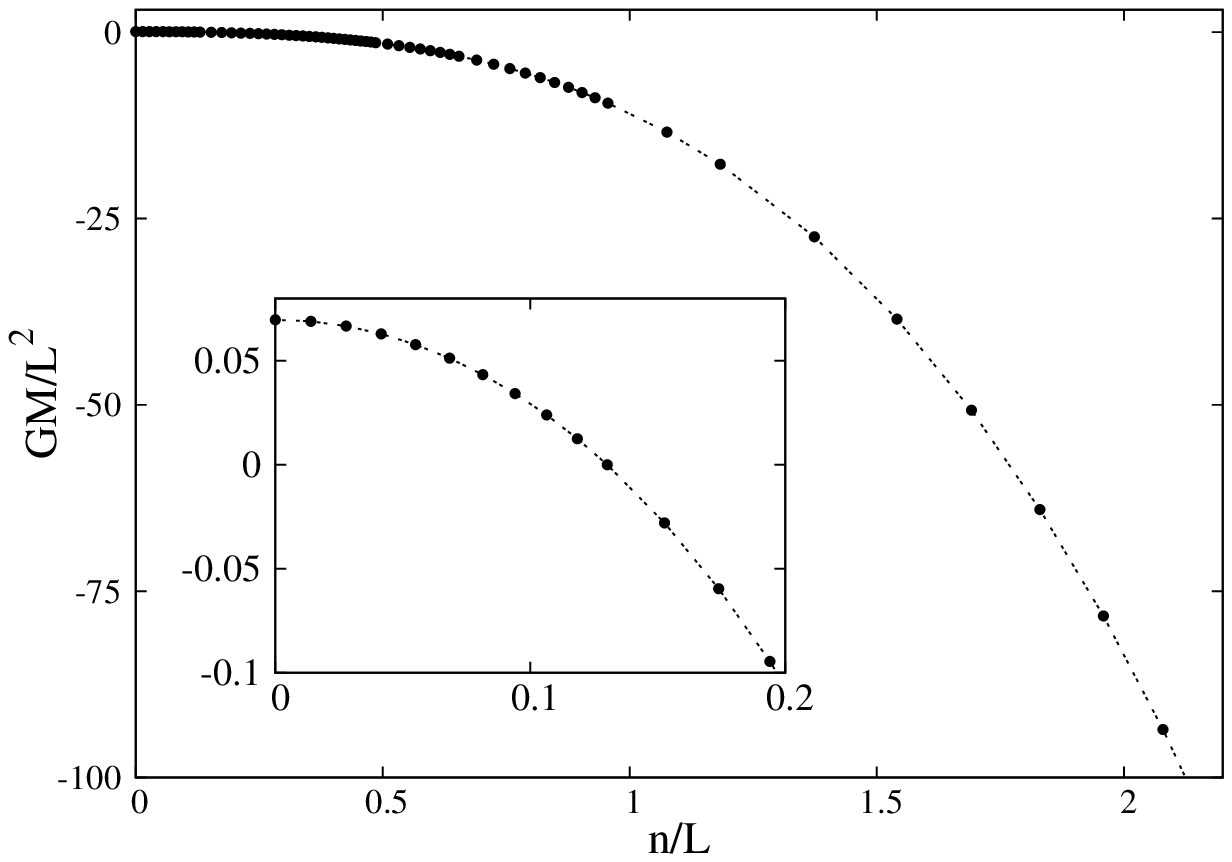,width=8cm}}
\put(8,0.0){\epsfig{file=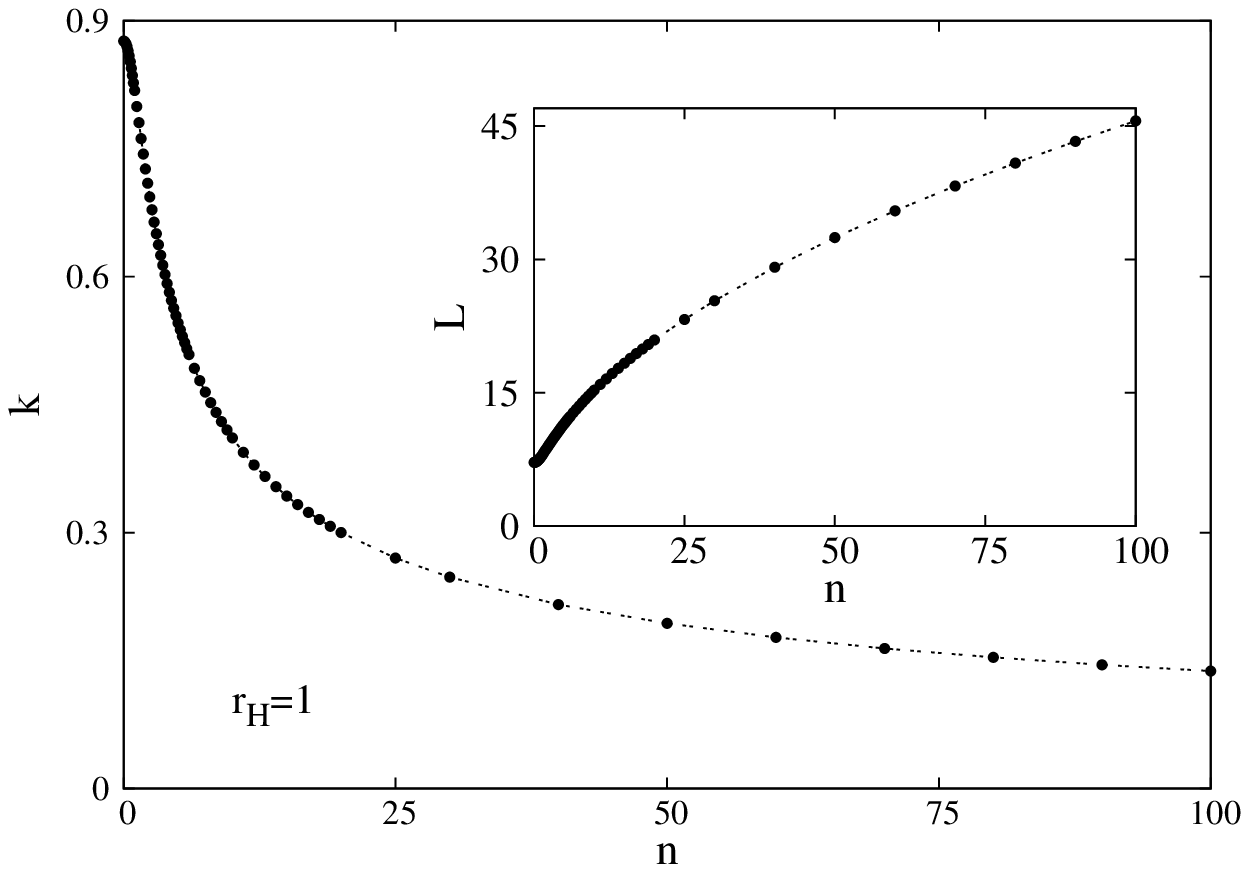,width=8cm}}
\end{picture}
\\
\\
{\small {\bf Figure 1.} 
{\it Left}: The critical mass $M$ is shown as a function of the NUT parameter $n$ 
(these quantities are given in units set by the length $L$
of the extra-dimension). 
{\it Right:} The wavelength $k=2\pi/L$ is shown as a function of the NUT parameter $n$
for solutions with a fixed horizon radius $r_H=1$
(the corresponding $L(n)$ curve is shown in the inset).
   }
\vspace{0.1cm}
%%%%%%%%%%%%%%%%%%%%%%%%%%%%%%%%%%%%%%%%%%%%%%%%%%%%%%%%%%%%%%%%%%%%%%%%%% 
\\
\\
where a prime denotes $d/d r$. 
This eigenvalue problem for the wavenumber $k=2\pi/L$ was solved numerically
with suitable boundary conditions.
First, the perturbation has to vanish for $r \to \infty$, 
$i.e.$
 ${\rm lim}_{r\to \infty} A_1,C_1 = 0$.
The solutions of the linearized equations should also be regular at the horizon.
This leads to a set of specific relations to be satisfied by $A_1(r_H)$, $C_1(r_H)$ and their derivatives.

To integrate the equations (\ref{eqs-p}), we have used the differential
equation solver COLSYS which involves a Newton-Raphson method
\cite{COLSYS}.
 In practice, we have set $r_H=1$ without loss of generality and
 computed the corresponding $k$ for a given value of $n$.
%Numerical solutions of (\ref{eqs-p}) were constructed in a 
%systematic way  for $0\leq n<100$
%and thus are likely to exist for any $n$.
  
Our numerical results show that 
the NUT-charged configurations inherit the GL instability of the purely ``electric" Schwarzschild black strings. 
Interestingly, for a given $L$,
the critical value of the mass descreases as $n$ increases, and becomes zero for 
$n/L \simeq 0.13021$.
The strings with $M<0$ become also unstable  for larger values of $\mu_2$.
The numerical results are displayed in Figure 1, where we exhibit the dimensionless quantity
$ \mu_1= M G/{L^{2}}$
$vs.$ the  dimensionless ratio between the NUT parameter and the length of the extra-dimension
$\mu_2= {n}/{L}$.
For completeness,  there we show also the value of $k$ as a function of $n$,
for a fixed value $r_H=1$ of the horizon radius
(the $L(n)$ curve is shown in the inset).

%%%%%%%%%%%%%%%%%%%%%%%%%%%%%%%%%%%%%%%%%%%%%%%%%%%%%%%%%%%%%%%%%%
\section{ The non-uniform solutions } 
%%%%%%%%%%%%%%%%%%%%%%%%%%%%%%%%%%%%%%%%%%%%%%%%%%%%%%%%%%%%%%%%%%

%%%%%%%%%%%%%%%%%%%%%%%%%%%%%%%%%%%%%%%%%%%%%%%%%%%%%%%%%%%%%%%%%%
\subsection{General relations}
%%%%%%%%%%%%%%%%%%%%%%%%%%%%%%%%%%%%%%%%%%%%%%%%%%%%%%%%%%%%%%%%%%  
As usual, the unstable GL mode signals the existence of 
a branch of solutions with a nontrivial
dependence on the extra-coordinate $z$.
These solutions are constructed numerically by using a 
similar approach to that employed in  
\cite{Kleihaus:2006ee},
\cite{Kleihaus:2007dg}
to construct NUBS
with usual Kaluza-Klein asymptotics.

The Einstein equations $G_t^t=0,~G_r^r+G_z^z=0$ and $G_\theta^\theta=0$  
 yield for the functions $A,~B,~C$ the set 
 of equations\footnote{
Note that the Einstein equations $G_z^r =0,~G_r^r-G_z^z  =0$ are not automatically satisfied,
yielding two constraints. However, following  \cite{Wiseman:2002zc}, one can show that
these constraints are satisfied as a consequence of the Bianchi identities.
Also, one can show that all other Einstein equations  are either linear combinations
of those used to derive 
(\ref{Eeq})-(\ref{Eeq2}) 
or are identically zero.
}
\begin{eqnarray} 
\label{Eeq}
&&A''+\frac{{\ddot A}}{f}+A'^2+\frac{{\dot A}^2}{f}
+ \frac{2r A'}{g}
+ \frac{3f' A'}{2f}
+\frac{ f' C'}{f}
+2(A'C'+\frac{{\dot A}{\dot C}}{ f})
+\frac{ f'r}{f g}
+\frac{f''}{2f}
+\frac{2e^{2A+2B-4C}n^2}{g^2}
=0,~~{~~~}
\end{eqnarray}
\begin{eqnarray}
\label{Eeq1}
&&B''+\frac{{\ddot B}}{f} 
-2( A'C'+\frac{{\dot A}{\dot C}}{ f}) 
-C'^2-\frac{{\dot C}^2}{f}
-\frac{2r A'}{g}
+\frac{f'B'}{2f}
-\frac{(2rf+g f')}{fg}C'
\\
\nonumber
&&{~~~~~~~~~~~~~~~~~~~~~~~~~~~~~~~~~~~~~~~~~~~~~~~~~~~~~~~~}
+\frac{1}{fg}(e^{2B-2C}-rf')
-\frac{r^2}{g^2}
+\frac{e^{2A+2B-4C}n^2}{g^2}
=0,
\end{eqnarray}
\begin{eqnarray}
\label{Eeq2}
&&C''+\frac{{\ddot C}}{f}
+2\left( C'^2+\frac{{\dot C}^2}{f} \right)
+A'C'+\frac{{\dot A}{\dot C}}{f}
+\frac{r A'}{g}
+\frac{f'C'}{f}
+\frac{4rC'}{g}
\\
\nonumber
&&{~~~~~~~~~~~~~~~~~~~~~~~~~~~~~~~~~~~~~~~~~~~~~~~~~~~~~}
+\frac{1}{fg}(rf'-e^{2B-2C})
+\frac{1}{g} -\frac{2e^{2A+2B-4C}n^2}{g^2}
=0,
\end{eqnarray}
where a prime denotes $\partial/\partial r$, and a 
dot $\partial/\partial z$. 
(Note that the $D=5$ equations in  Ref. \cite{Kleihaus:2006ee} are recovered for $n=0$.)

To solve these equations, we use the same approach as for $n=0$,
and introduce a new radial coordinate $\tilde r$, where $r=\sqrt{r_H^2 + \tilde r^2}$ 
($i.e.$ the horizon resides at $\tilde r=0$). 
Utilizing the reflection symmetry of the solutions 
w.r.t.~$z=L/2$,
the solutions are constructed subject to the 
boundary conditions
\begin{eqnarray}
\label{bc1} 
\partial_z A\big|_{z=0,L/2}=\partial_z B\big|_{z=0,L/2}
=\partial_z C\big|_{z=0,L/2}=0,
~~
A\big|_{\tilde r=0}-B\big|_{\tilde r=0}=d_0,~\partial_{\tilde r} 
A\big|_{\tilde r=0}=\partial_{\tilde r} C\big|_{\tilde r=0}=0,~ 
\end{eqnarray}
(where the constant $d_0$ is related to the Hawking 
temperature of the solutions (\ref{temp})),
together with
\begin{eqnarray}
%&&
%\nonumber
\label{bc3} 
A\big|_{\tilde r=\infty}=B\big|_{\tilde r=\infty}=C\big|_{\tilde r=\infty}=0,~~
\end{eqnarray}
such that the uniform background $NUT\times S^1$ is approached asymptotically. 
Regularity further requires that the condition 
$\partial_{\tilde r} B\big|_{\tilde r=0}=0$ holds for the solutions.
  
The asymptotic form of the  relevant metric components is
\begin{eqnarray}
\label{1} 
g_{tt}\simeq -1+\frac{c_t}{r},~~~g_{zz}\simeq 1+\frac{c_z}{r},
\end{eqnarray} 
and contains two parameters $c_t$ and $c_z$
encoding the global charges of the solutions 
(with $c_t=2m$, $c_z=0$ for uniform configurations).

The global charges of the NUT strings 
are the  mass $M$ and the tension $T$.
In their computation,
it is convenient to use again
the quasilocal formalism augmented by the counterterm approach\footnote{
For 
$D=5$  solutions, the appropriate expression of the counterterm is 
$I_{ct}=-\frac{1}{8 \pi G}\int_{\partial {\cal M}} d^{4} x \sqrt{-h}\sqrt{2 \mathcal{R}}$.}.
$M$ and $T$ are charges associated with the asymptotic Killing vectors $\partial/\partial t$ and  $\partial/\partial z$, 
respectively,
their expressions being  analogous to those valid in the $n=0$ limit, with\footnote{Note that
the non-uniform solutions possess also a nonzero angular momentum density.
However, similar to the case of  the $D=4$ NUT solution,
the total angular momentum vanishes.}
\begin{eqnarray}
\label{2} 
M=\frac{ L}{4 G}(2c_t-c_z),
~~
T=\frac{1}{4\pi G}(c_t-2c_z).
\end{eqnarray}  

Other quantities of interest are 
the Hawking temperature and the horizon area of the non-uniform solutions
\begin{eqnarray}
\label{temp} 
T_H=\frac{1}{4 \pi r_H}e^{A_0-B_0},~~~A_H= 4\pi L  (r_H^{2}+n^2) \int_0^L e^{B_0+2C_0}dz,
\end{eqnarray} 
where $A_0(z),B_0(z),C_0(z)$ are the values of the metric functions on the event horizon $r=r_H$.

To obtain a 
  measure of the deformation of the
solutions,
we define the non-uniformity parameter \cite{Gubser:2001ac}
\begin{equation} 
\lambda = 
\frac{1}{2} 
\left( \frac{{\cal R}_{\rm max}}{{\cal R}_{\rm min}}
 -1 
 \right), 
\label{lambda} 
\end{equation}

%%%%%%%%%%%%%%%%%%%%%%%%%%%%%%%%%%%%%%%%%%%%%%%%%%%%%%%%%%%%%%%%%%%%%%%%%%%%%% 
\setlength{\unitlength}{1cm}
\begin{picture}(8,6) 
\put(-0.5,0.0){\epsfig{file=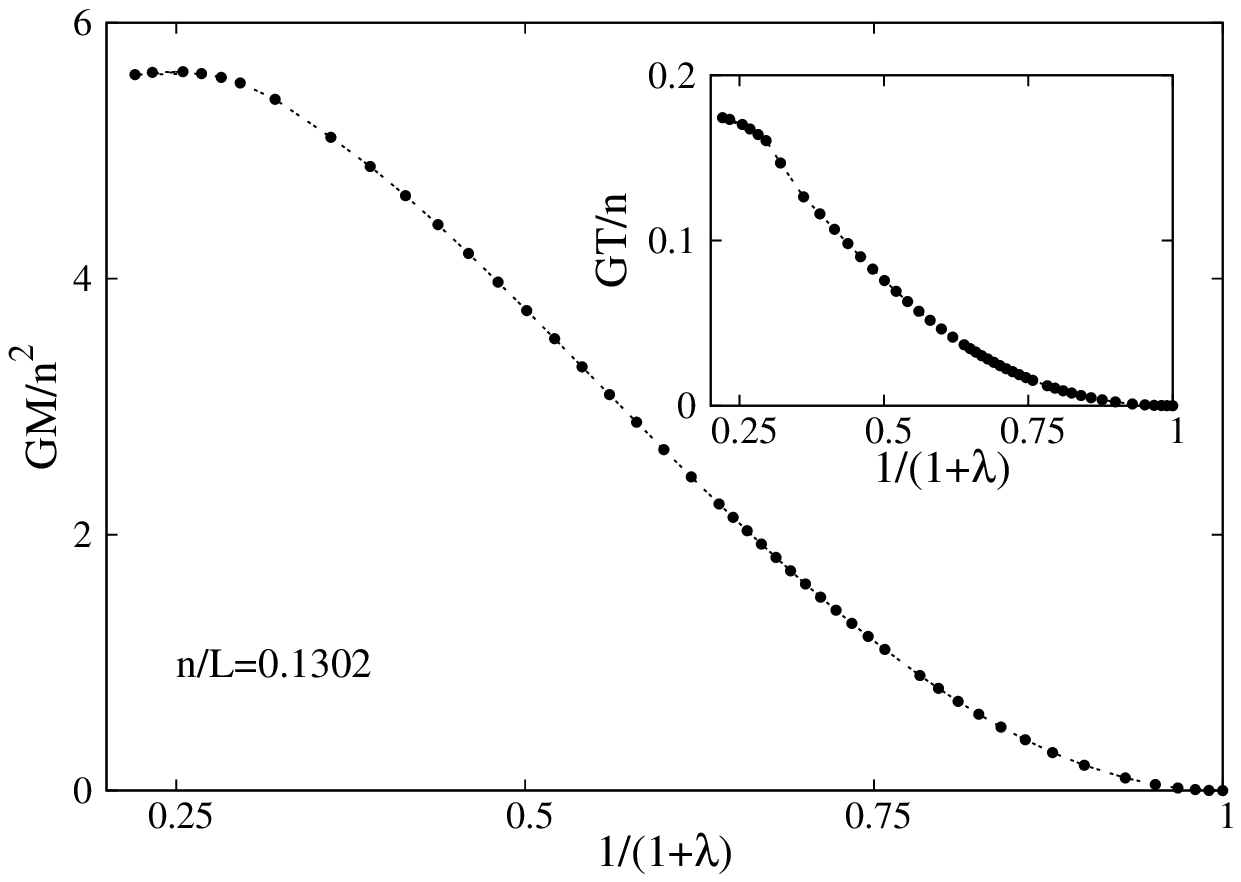,width=8cm}}
\put(8,0.0){\epsfig{file=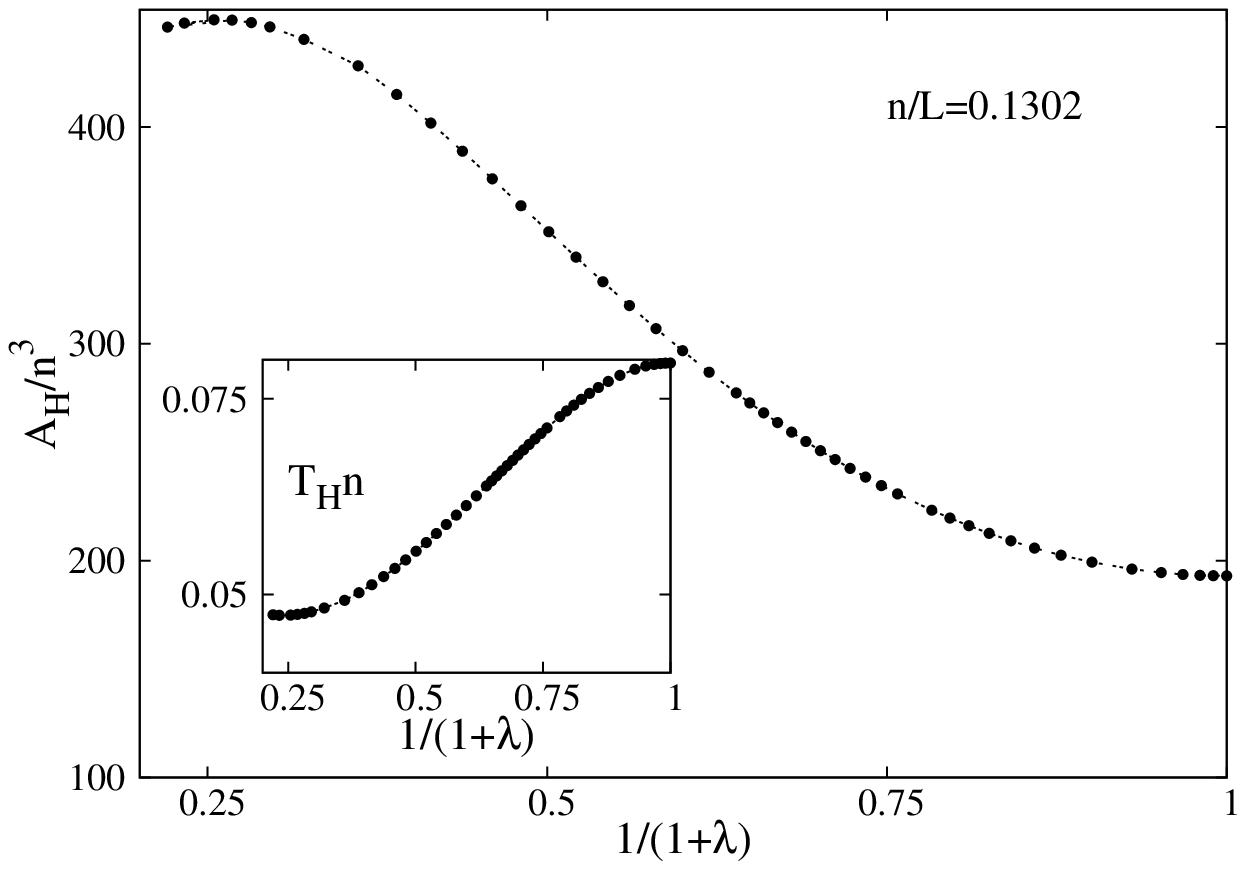,width=8cm}}
\end{picture}
\\
\\
{\small {\bf Figure 2.} 
The mass $M$, 
the  tension $T$, 
the temperature $T_H$ 
and the area $A_H$ 
are shown in units of $n$  as functions of  the
non-uniformity 
$1/(1 + \lambda)$,
for a family of
 non-uniform strings with
$\mu_2=n/L=0.1302$.
   }
\vspace{0.1cm}
%%%%%%%%%%%%%%%%%%%%%%%%%%%%%%%%%%%%%%%%%%%%%%%%%%%%%%%%%%%%%%%%%%%%%%%%%%
\\
\\ 
 where ${\cal R}_{\rm max}$ and ${\cal R}_{\rm min}$
represent the maximum and minimum radii of the two-sphere on the horizon.
 
We remark also that the  equations (\ref{Eeq})-(\ref{Eeq2}) are left invariant
by the transformation 
$r \to r /p$, 
$ z \to z /p$, 
$r_H \to r_H/ p$,
$n \to n /p$,
with $p$ a positive integer.
Therefore, a new family of vacuum solutions
with the same length of the extradimension 
can be generated 
in this way 
(see $e.g.$ \cite{Harmark:2003eg}
for a detailed discussion of this procedure 
for the $n=0$ case).
Their relevant quantities, expressed in terms of 
  the initial solution, read
%\begin{eqnarray}
$M^{(p)}={M}/{p},$
$T^{(p)}= {T}/{p},$
$T_H^{(p)}=p T_H,$
$A^{(p)}=A_H/{p^2}$.
%\end{eqnarray}

%%%%%%%%%%%%%%%%%%%%%%%%%%%%%%%%%%%%%%%%%%%%%%%%%%%%%%%%%%%%%%%%%%
\subsection{The solutions}
%%%%%%%%%%%%%%%%%%%%%%%%%%%%%%%%%%%%%%%%%%%%%%%%%%%%%%%%%%%%%%%%%% 

The  
non-linear elliptic partial differential equations (\ref{Eeq})-(\ref{Eeq2}) are solved
numerically, subject to the boundary conditions  (\ref{bc1}),  (\ref{bc3}).  
The numerical calculations
are based on the Newton-Raphson method and are performed 
with help of the professional package FIDISOL/CADSOL \cite{schoen},
 which provides also an error estimate for each unknown function.
(See the Ref. \cite{Kleihaus:2006ee}
 for a detailed description of the numerical scheme.)
 
The input parameters of the problem are 
the horizon coordinate $r_H$, 
the temperature $T_H$,
the NUT charge $n$
and the asymptotic length $L$ of the compact $z-$direction.

For a given $n$,
a branch of non-uniform solutions is obtained by starting at the critical point of the uniform
configurations and varying the boundary parameter $d_0$, 
which enters  Eq. (\ref{bc1}), relating
the values of the functions $A$ and $B$ at the horizon.
Our numerical results show that the properties of the solutions are 
rather similar to the case of the non-uniform generalizations of the $n=0$
Schwarzschild black string.
In particular, the functions $A,B,C$ 
have a similar shape to that displayed in Ref. \cite{Kleihaus:2006ee},
%The functions $A,B,C$ 
exhibiting extrema
at $z = 0$ at the horizon. 
As $\lambda$ increases, the extrema increase in height
and become increasingly sharp.

Some numerical results are displayed in Figure 2, 
where we exhibit 
 the mass $M$, 
 the  tension $T$, 
the temperature $T_H$ and the
horizon area $A_H$  
 versus the parameter $\lambda$ for a family of non-uniform solutions
 with $\mu_2=n/L=0.1302$.
In that plot, $M$, $T$, $T_H$ and $A_H$ 
 are given in units of $n$, with $\lambda=0$ corresponding to the uniform solution.
One can see that
the mass and horizon area
 assume a maximal value for a value of $\lambda=\lambda_{ex}$,
 where the temperature assumes a minimal value.

Non-uniform  strings can also be obtained by starting from $n=0$ NUBSs
with a given temperature
(as specified by the parameter $d_0$) and length of the extra-dimension,
and then slowly increasing the value of  the NUT charge.
The numerical results suggest that any  NUBS possess  
generalizations with $n\neq 0$.
No upper bound on $n$ appears to exist, although the numerical
integration becomes more difficult with increasing $n$.

%%%%%%%%%%%%%%%%%%%%%%%%%%%%%%%%%%%%%%%%%%%%%%%%%%%%%%%%%%%%%%%%%%%%%%%%%%%%%% 
\setlength{\unitlength}{1cm}
\begin{picture}(8,6) 
\put(-0.5,0.0){\epsfig{file=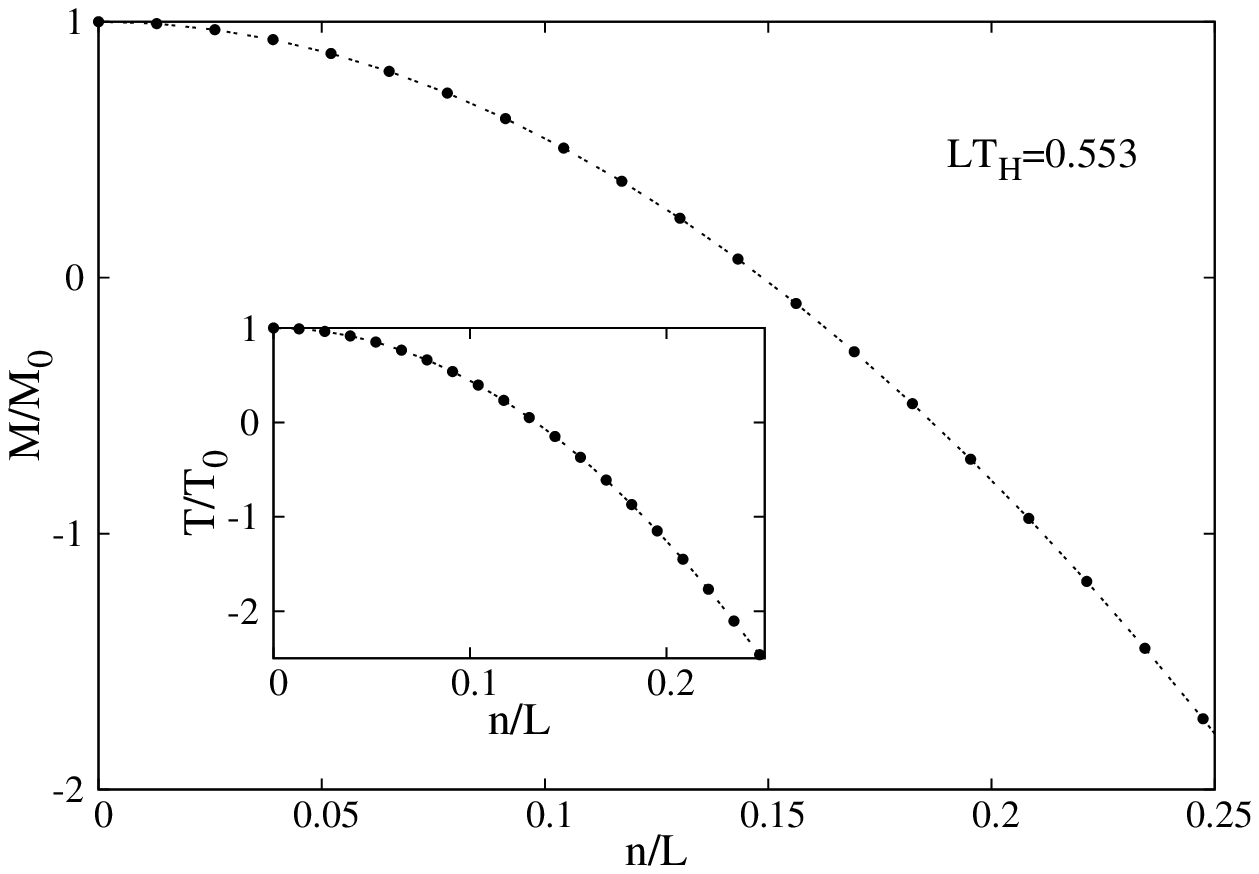,width=8cm}}
\put(8,0.0){\epsfig{file=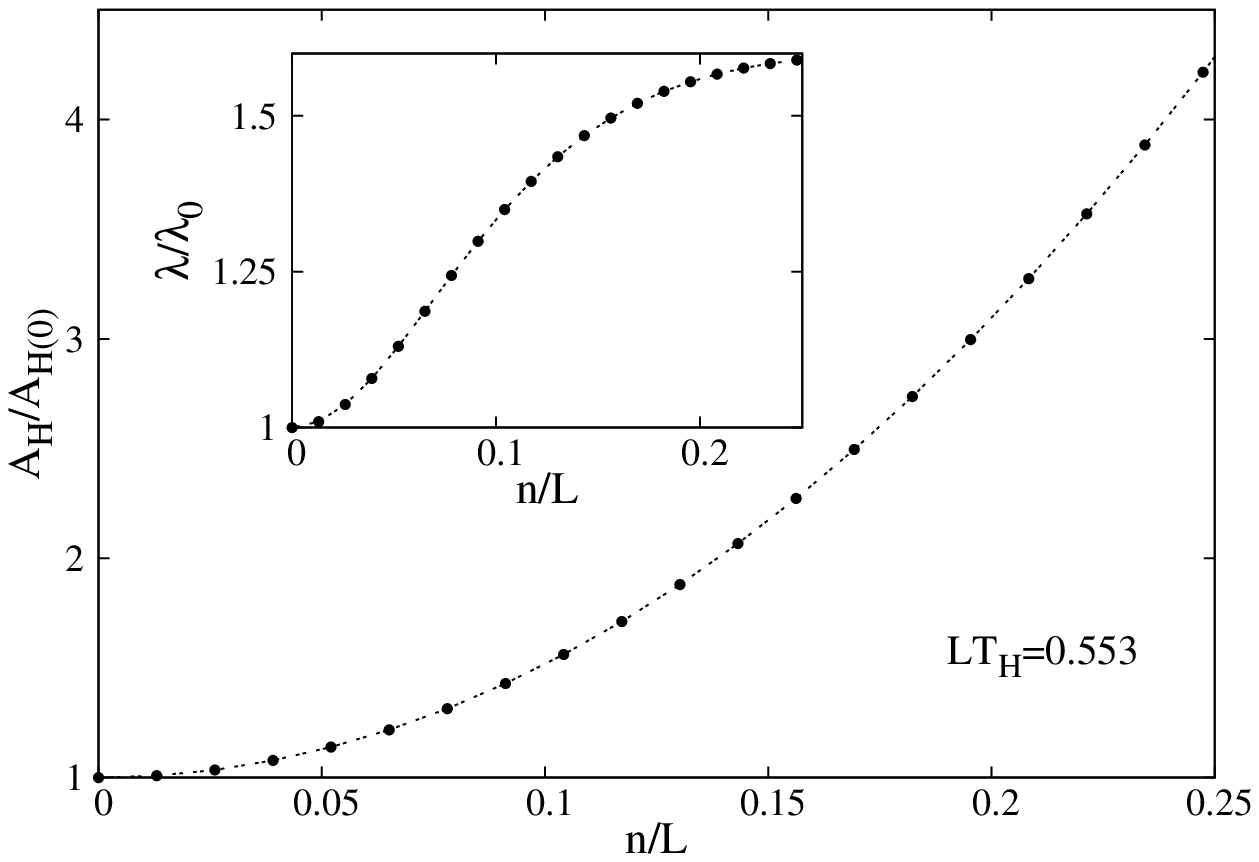,width=8cm}}
\end{picture}
\\
\\
{\small {\bf Figure 3.} 
The mass $M$, 
the  tension $T$,
the area $A_H$
and
the non-uniformity parameter $\lambda$
of a family of solutions with $L T_H=0.553$
are shown in units of the corresponding $n=0$ NUBS  
solution 
(denoted by $M_0$, $T_0$, $A_{H(0)}$ and $\lambda_0$),
 as functions of
 the dimensionless ratio $n/L$. 
   }
\vspace{0.1cm}
%%%%%%%%%%%%%%%%%%%%%%%%%%%%%%%%%%%%%%%%%%%%%%%%%%%%%%%%%%%%
\\
\\
In Figure 3 we exhibit the mass $M$, tension $T$, horizon area $A_H$
and non-uniformity parameter $\lambda$
for 
non-uniform string solutions with $L T_H=0.553$, 
in units of the corresponding
$n=0$ solution, versus the parameter $\mu_2=n/L$.
As one can see, both the mass and tension decrease with $n$,
becoming negative for large enough values of  the NUT charge.
At the same time,  the horizon area
and the non-uniformity parameter increase.

%%%%%%%%%%%%%%%%%%%%%%%%%%%%%%%%%%%%%%%%%%%%%%%%%%%%%%%%%%%%%%%%%%
\section{Further remarks}
%%%%%%%%%%%%%%%%%%%%%%%%%%%%%%%%%%%%%%%%%%%%%%%%%%%%%%%%%%%%%%%%%%

The main purpose of this work was to investigate the 
stability of the Lorentzian 
NUT solution uplifted to $D=5$ dimensions.
Even if one's primary interest is in solutions with usual Kaluza-Klein asymptotics, 
we hope that, by widening the
context to solutions with NUT charge, one may achieve a deeper appreciation of the theory.  
In particular, one may hope to determine more general features of the string solutions, 
independent of
whether or not they contain a ``magnetic" mass. 
Expanding around the uniform solution
and solving the eigenvalue problem numerically, our results indicate that the GL instability
persists for $n\neq 0$ configurations.
Moreover, for a given length of the extra-dimension, a $D=5$ NUT-charged solution
becomes unstable for a smaller value of the mass as compared to 
the Schwarzschild black string.
 
We also constructed numerically the corresponding non-uniform strings 
emerging from the branch of marginally stable uniform solutions.
The properties of these solutions are rather similar to the well-known
$n=0$ case.
An interesting point which remains to be clarified is the 
phase diagram of the $D=5$
solutions approaching at infinity
a $NUT\times S^1$ background.
For $n=0$, 
apart from the black string solutions, the Kaluza-Klein theory possesses also a branch of black hole
solutions with an  $S^3$ topology of the event horizon.
There is now convincing evidence that the non-uniform string branch and the black
hole branch merge at a topology changing solution.
Based on the numerical results 
in Section 4,
we expect that a similar picture should be valid  
also for the configurations in this work.
The conjectured horizon topology changing transition should be approached again for
$\lambda \to \infty$.
However, the construction of the $D=5$ nutty solutions with an $S^3$ horizon topology
still represents a  numerical challenge.
 
Finally,
let us remark that the NUT solution (\ref{TN})
possesses higher dimensional  $D=2K+2$ generalizations
(see $e.g.$ \cite{Mann:2005ra} and the references there).
Their main properties (in particular the presence of closed timelike curves) 
follow closely the four-dimensional case.
These generalized NUT solutions can also be uplifted to $D=2K+3$ dimensions
and are likely to possess as
well a GL unstable mode.

\vspace*{0.5cm}

%%%%%%%%%%%%%%%%%%%%%%%%%%%%%%%%%%%%%%%%%%%%%%%%%%%%%%%%%%%%%%%%%%%%%%%%%%%%%%%%%%
\noindent{\textbf{~~~Acknowledgements.--~}}  
We gratefully acknowledge support by the DFG,
in particular, also within the DFG Research
Training Group 1620 ``Models of Gravity''. 
 %%%%%%%%%%%%%%%%%%%%%%%%%%%%%%%%%%%%%%%%%%%%%%%%%%%%%%%%%%%%%%%%%%%%%%%%%%%%%%%%%%%%%  

%%%%%%%%%%%%%%%%%%%%%%%%%%%%%%%%%%%%%%%%%%%%%%%%%%%%%%%%%%%%%%%%%%%%%%%%%%%%%%%%%%%%%  
 \begin{small}
 
%%%%%%%%%%%%%%%%%%%%%%%%%%%%%%%%%%%%%%%%%%%%%%%%%%%%%%%%%%%%%%%%%%%%%%%%%%%%%%
 \end{small}

 \end{document}